# Colloidal quasi-2D Cs$_2$AgBiBr$_6$ double perovskite nanosheets: synthesis and application as high-performance photodetectors

Pannan I. Kyesmen[1], Eugen Klein[1], Brindhu Malani S[1], Rostyslav Lesyuk[1], and Christian Klinke[*1,2]

[1]Institute of Physics, University of Rostock, Albert-Einstein-Str. 23-24, 18059 Rostock, Germany
[2]Department Life, Light & Matter, University of Rostock, Albert-Einstein-Strasse 25, 18059 Rostock, Germany

* Correspondence should be addressed: Christian.klinke@uni–rostock.de

**Abstract**

The search for non-toxic lead-free halide perovskites that can compete with the lead-based counterparts has led to the emergence of double perovskites as potential candidates. Among many options, Cs$_2$AgBiBr$_6$ stands out as one of the most suitable eco-friendly materials for numerous optoelectronic applications. In this study, quasi-2D Cs$_2$AgBiBr$_6$ nanosheets (NSs) were prepared via the low-temperature injection colloidal synthesis and used to fabricate high-performance photodetectors in a transport-layer-free architecture. The reaction temperature and ligands played vital roles in the structural purity, shape, and size of the synthesized Cs$_2$AgBiBr$_6$ NSs. The fabricated NSs disclosed lateral sizes of up to 1.4 μm and are only a few nanometers thick. The high-performance photodetectors fabricated using the Cs$_2$AgBiBr$_6$ NSs yielded a high detectivity (D) of $1.15 \times 10^{12}$ Jones, responsivity (R) of 121 mA/W, a notable on/off ratio of $2.39 \times 10^4$, and a fast rise and decay time of 857 and 829 μs, respectively. The device demonstrates remarkable stability. Basically, it sustains its entire photocurrent after storage in ambient conditions for 80 days. This work showcases a pathway for the colloidal synthesis of quasi-2D Cs$_2$AgBiBr$_6$ lead-free double perovskite NSs with suitable properties for high-performance photodetection and other optoelectronic applications.

## 1.0 Introduction

The difficulty in managing lead (Pb) toxicity and stability issues in lead halide perovskites has steered an intense search for stable and lead-free perovskites that can yield comparable properties to their lead-based counterparts for many optoelectronic applications. The earlier solutions attempted to directly replace Pb in the $APb^{2+}X_3$ basic composition with a similar divalent cation such as Sn and Ge. For Sn, the huge number of defects and stability issues arising from the easy oxidation of $Sn^{2+}$ to $Sn^{4+}$ under ambient conditions poses a major challenge for their application in optoelectronic devices [1, 2]. On the other hand, a few experimental studies have been reported for Ge-based perovskites [3-5]. However, Ge is a rare metal, and its scarcity makes it very expensive and unattractive for practical applications in photovoltaic and optoelectronic devices [6]. Moreover, it also suffers from stability issues and can easily oxidize from $Ge^{2+}$ to $Ge^{4+}$ [7]. The emergence of double perovskites has opened promising options for developing low-cost and environmentally friendly halide perovskites for photovoltaic and optoelectronic applications. The heterovalent replacement of two divalent $Pb^{2+}$ ions in Pb-based perovskites with one monovalent ion ($Na^+$, $Ag^+$, $Cu^+$) and one trivalent ion ($Bi^{3+}$, $In^{3+}$, $Sb^{3+}$) forms a diverse class of promising quaternary $A_2M^+M^{3+}X_6$ double perovskites. This configuration ensures charge neutrality similar to conventional lead-based perovskites, structural diversity, and controllable size, shape, and electronic band properties for a more stable material [8].

Many Pb-free double perovskites, including $Cs_2AgInCl_6$ [9], $Na_2CuBiBr_6$ [10], and $Cs_2NaBiCl_6$ [11], have been developed after the first reports of $Cs_2AgBiBr_6$ bulk crystals in early 2016 [12, 13]. Despite these options, $Cs_2AgBiBr_6$ is still one of the most suitable eco-friendly materials for numerous optoelectronic applications. The material possesses desirable properties such as long carrier lifetimes, high chemical, thermal, and ambient stability, relatively lower effective mass of charge carriers, insignificant toxicity, and high light absorption coefficient [14, 15], beneficial for various optoelectronic applications. When prepared as confined 2D/quasi-2D structures, they offer the unique advantage of quantum confinement effects and wide bandgap tunability, which can be relevant for effective engagement in applications such as photovoltaics, photoelectrochemistry, and photodetection.

Photodetectors based on 2D/quasi-2D materials are among the most competitive candidates for efficient and cost-effective device design [16]. This is because 2D/quasi-2D materials are typically synthesized in their single-crystalline states and have fewer ionic defects and grain boundaries when compared to their corresponding 3D polycrystalline films [17]. Moreover, when in confinement, 2D/quasi-2D materials exhibit intense absorbance at specific wavelengths that permit optical tunability and wavelength-selective photodetection applications. 3D materials of $Cs_2AgBiBr_6$ double perovskites have exhibited promising photodetection properties [18-20]. However, the presence of unavoidable defects and a huge level of grain boundaries in 3D polycrystalline particles results in large photocurrent hysteresis, reduced photostability and moisture tolerance, and decreased performance reliability [17, 21]. Meanwhile, 2D $Cs_2AgBiBr_6$ double perovskites exhibit low intrinsic carrier density and high resistivity in the out-of-plane direction, which leads to low dark current and consequently high detectivity. This has been demonstrated in a previous study where 2D $Cs_2AgBiBr_6$ fabricated using the space-confined technique yields superior photodetection properties over their 3D counterparts [15]. Typical 2D perovskites form bound excitons with a higher binding energy when excited, which limits conductivity (in the out-of-plane direction) and charge transport in photodetection applications [21]. However, quasi-2D perovskites can harness the benefits of both 2D and 3D structures for better device design and performance, which has been demonstrated in many applications, including photovoltaic [22], supercapacitors [23], and photodetection [24]. The quantum confinement effects, high surface-to-volume ratio, high absorption coefficient, and out-of-plane high resistivity make 2D/quasi-2D $Cs_2AgBiBr_6$ attractive for efficient photodetection design for practical applications.

The ball milling approach [25], space-confined fabrication [15], and colloidal synthesis [26] are among the techniques that have been engaged for the preparation of confined 2D/quasi-2D nanostructures of $Cs_2AgBiBr_6$. The colloidal synthesis offers a versatile, low-cost approach for preparing high-quality, quantum-confined single crystals of $Cs_2AgBiBr_6$. In 2021, Liu *et al.* developed a low-temperature injection colloidal technique for the first synthesis of quantum-confined 2D $Cs_2AgBiBr_6$ double perovskite nanoplatelets [27]. Further work done in 2023 by Dor et al. provides additional insight into the growth mechanism of the nanoplatelets and proposed self-assembly of small platelets into larger ones during the crystal formation, which results in some physical defects [28]. In both cases, the confined nanoplatelets were prepared at a reaction temperature of 230 °C, and consisted of platelets with a lateral size of 330 ± 230 nm and a thickness of 3.6-6.0 nm. These studies did not provide a comprehensive report on the role of reaction temperature and ligands in the growth of colloidal $Cs_2AgBiBr_6$ nanocrystals. More so, the successful colloidal synthesis of $Cs_2AgBiBr_6$ nanoplatelets/nanosheets above 230 °C has not been reported based on known literature. In colloidal synthesis, the reaction temperature and ligands can significantly influence the size, shape, structural defects, and crystallization of nanomaterials. In the colloidal synthesis of perovskite nanocrystals, an increase in the concentration of ligands can provide a unique acid-base equilibrium that could allow their crystallization at a higher temperature, thereby influencing their size, shape, and structure [29].

In this project, colloidal quasi-2D $Cs_2AgBiBr_6$ nanosheets (NSs) were fabricated using a low-temperature injection and heat-up process, and the role of reaction temperature and ligands in the synthesis was systematically investigated. The $Cs_2AgBiBr_6$ NSs were used to develop high-performance photodetectors. Quasi-2D $Cs_2AgBiBr_6$ NSs with a lateral size of up to 1.4 μm and only a few nm thick were fabricated at 250 °C reaction temperature under favorable ligand conditions, without distorting the material's crystal structure. The reaction temperature and ligands served crucial roles in the structural integrity, shape, and size of the synthesized $Cs_2AgBiBr_6$. The high-performance photodetectors show a high detectivity of $1.15 \times 10^{12}$ Jones, responsivity of 121 mA/W, and fast rise and fall times of 857 and 829 μs, respectively. In addition, the device demonstrated a notably high on/off ratio of $2.39 \times 10^4$. It is worth noting that the $Cs_2AgBiBr_6$ photodetectors attained such high performance in a charge transport layer-free architecture, which usually adds to the complexity of device design and cost effectiveness. This work paves a pathway for developing large quantum-confined NSs of colloidal Pb-free $Cs_2AgBiBr_6$ double perovskites with suitable properties for high-performance photodetection and other optoelectronic applications.

## 2.0 Results and discussion

### 2.1 Cs$_2$AgBiBr$_6$ Nanosheets

Nanosheets of Cs$_2$AgBiBr$_6$ double perovskites were synthesized using the low-temperature crystallization approach first developed by Liu et al. in 2021, with some modifications to the final reaction temperature and amount of oleic acid (OA) ligand used in the synthesis [27]. An increase in the amount of OA (from 1.0 ml to 1.5 ml) allowed for the synthesis of Cs$_2$AgBiBr$_6$ NSs at a higher final reaction temperature of 250 °C (instead of 230 °C), allowing for the synthesis of laterally larger NSs without any structural distortion. In brief, the precursors; silver nitrate (AgNO$_3$), bismuth bromide (BiBr$_3$), 1-octadecene (ODE), OA, oleylamine (OLAm), and hydrobromic acid (HBr) were added into a three-necked flask, degassed at 120 °C, and heated to 200 °C to fully dissolve the mixture under continuous Ar flow. The solution was cooled to room temperature, followed by the injection of Cs-oleate and stirring for 10 min. The mixture was then heated to 250 °C for 10 min to obtain NSs of Cs$_2$AgBiBr$_6$. The growth mechanism of the nanostructures through the self-assembly of smaller platelets into larger ones has been proposed and explained in a previous study [28]. Given appropriate reaction conditions and a higher reaction temperature, the small nanoplatelets will acquire more thermal and kinetic energy, causing them to migrate faster, increase their reactivity, and yield larger NSs. The roles that ligands and temperature play in preparing quantum-confined NSs of Cs$_2$AgBiBr$_6$ double perovskites are investigated and will be discussed in more detail in section 2.2.

**Figure 1a** shows the transmission electron microscopy (TEM) image of the synthesized Cs$_2$AgBiBr$_6$ NSs (**Figure S1a** shows a more expanded view at a lower magnification). The micrograph revealed NSs with rectangular shapes and a wide lateral size distribution of 0.3-1.4 µm. The maximum lateral size of 1.4 µm observed for the NSs is way more than double the highest lateral dimension previously obtained for the colloidal nanoplatelets of the material [27, 28]. The lateral growth of the nanoplatelets into larger NSs was induced by an increase in the quantity of OA in the reaction solution, which provided a suitable acid-base equilibrium for the ligands and enabled the appropriate shelling of the platelets that allowed for the reaction to be sustained at 250 °C without distorting the structure and composition of the material. Higher reaction temperature increases the thermal energy in the system, which enhances the migration and growth of smaller platelets into NSs. For application of Cs$_2$AgBiBr$_6$ as photodetectors, NSs with higher lateral dimensions will effectively bridge the electrode spacing and reduce grain boundaries, which can serve as traps for charge carriers, thereby promoting charge transport and device performance [30, 31]. X-ray diffraction (XRD) measurements conducted on the fabricated NSs showed patterns that emphatically aligned with the cubic phase of the crystal structure of Cs$_2$AgBiBr$_6$ according to ICSD card number 01-087-9014 and exhibited an orientation in the <001> direction when deposited on the substrates, as seen in the high intensity observed for (400) and (200) planes (**Figure 1b**). The low-angle diffraction patterns revealed periodic peaks, which indicate that the NSs are stacked along the <001> direction (**Figure S1b**). The periodicities of 4.0 and 3.5 nm were deduced from equidistant reflections in the low-angle XRD patterns, which reveal a stacking behavior of single ultrathin nanosheet layers to form the quasi-2D nanostructures. The difference between these two values is approximately equal to a half unit cell of the Cs$_2$AgBiBr$_6$ (~0.57 nm, called further as monolayer (ML)), meaning that the thickness of individual 2D layers increases by one ML of [Ag, Bi]Br octahedrons. The size of a unit cell was calculated from the XRD data using Bragg's law and the Miller indices (**Table S1**). Considering the length of OLAm to be ~1.7 nm [28], the thickness of the individual nanosheet layer is then estimated to be ~1.8 nm in the first case and ~2.3 nm in the second case. The high-resolution TEM (HR-TEM) performed on a single nanosheet of $Cs_2AgBiBr_6$ further affirms good crystallinity with clear lattice fringes over the whole sheet (**Figure 1c**). Lattice d-spacing of 2.8 and 4.0 Å were calculated from the lattice fringe and assigned to the (400) and (220) of the cubic structure of bulk $Cs_2AgBiBr_6$. The fast Fourier transform (FFT) pattern derived from the HR-TEM analysis further affirms the material's cubic structure (**Figure S2**). Selected area electron diffraction (SAED) analysis conducted on a single sheet also shows growth along the $[00\bar{1}]$ zone axis (**Figure 1d**) for the cubic crystal of $Cs_2AgBiBr_6$. The lattice spacing calculated from the observed planes in the SAED image matches well with the values calculated from the XRD pattern of the $Cs_2AgBiBr_6$ NSs, as given in Table S1. TEM/energy dispersive X-ray spectroscopy (TEM-EDS) measurements detected signals for all the elements of the NSs with an atomic weight ratio of 1.9:1.0:1.1:6.8 for Cs:Ag:Bi:Br in line with the expected 2:1:1:6 stoichiometry of the material (**Figure S3a**). EDS elemental mapping done on $Cs_2AgBiBr_6$ NSs (HR-TEM micrograph shown in **Figure S3b**) revealed a uniform distribution of the constituent elements across the material's surface (**Figure S3c-f**). Atomic force microscopy (AFM) analysis revealed stacked NSs that are 5.7-19.8 nm thick (**Figure 1e and f**). The thinnest structures of ~5.7 nm align well with the consideration of two OLAm layers with an inorganic part of ca. 2 nm. The thickness of individual ultra-thin layers would then correspond to about 1.5 unit cells or 3 [Ag, Bi]Br ML, and the thickness of 2.3 nm would correspond to 4 ML.

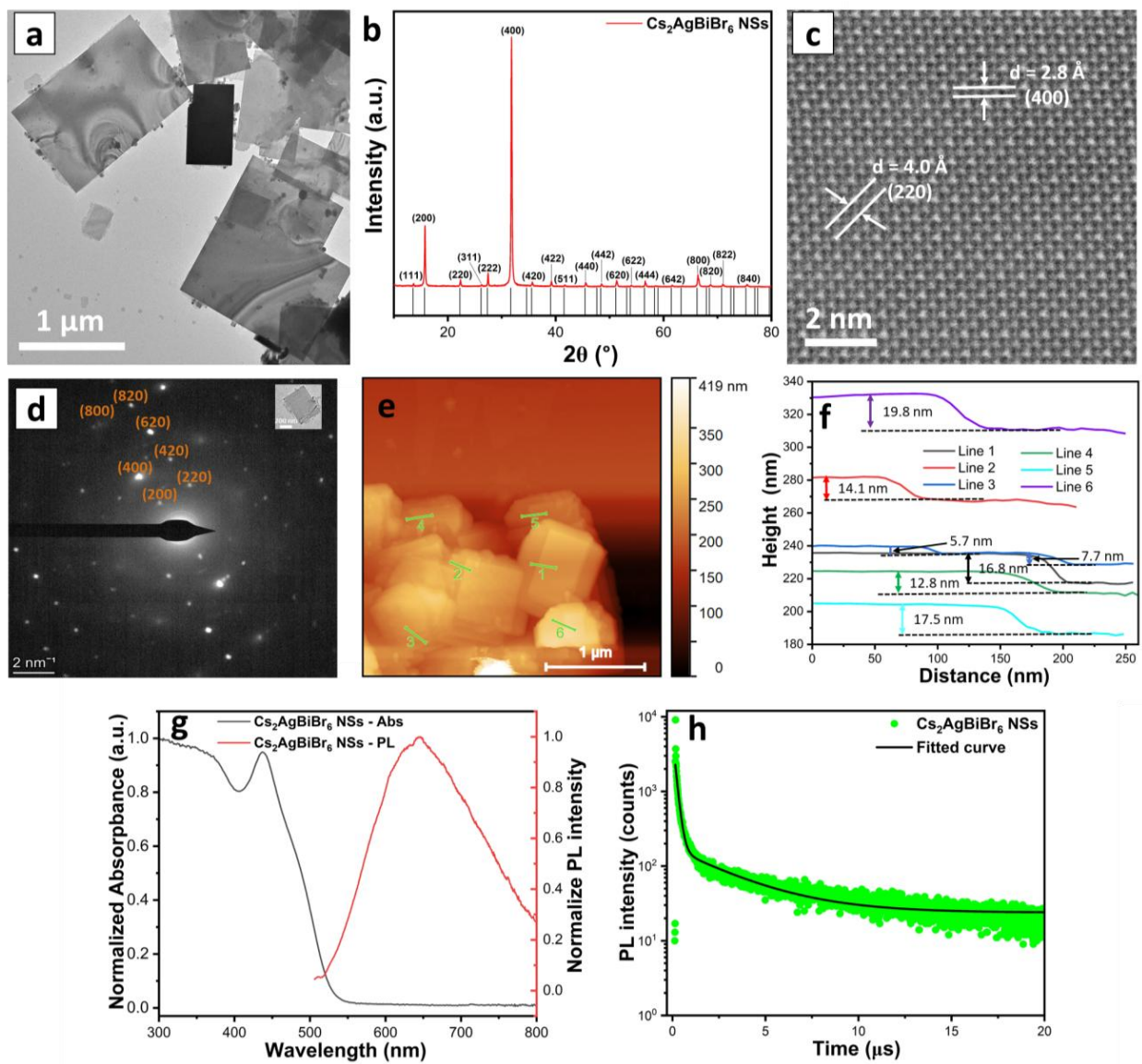

**Figure 1**. The synthesized Cs$_2$AgBiBr$_6$ NSs. a) TEM micrograph, b) XRD pattern (cubic phase of Cs$_2$AgBiBr$_6$ according to ICSD card number 01-087-9014 ), c) HRTEM image, d) SAED pattern, e) AFM image showing NSs and f) their height profiles, g) UV-Vis absorption and PL spectra, h) time-resolved PL decay and its corresponding fitted curve.

The absorption spectrum of the Cs$_2$AgBiBr$_6$ NSs is given in **Figure 1g** and revealed an excitonic absorption peak centered at 2.83 eV (437.8 nm), which is near the band edge, and is due to the 6s–6p transition of Bi in the [BiBr$_6$]$^{3-}$ octahedra of the material's crystal structure [32, 33]. The absorption at higher energies with a shoulder at about 3.40 eV (365.1 nm) results from direct bandgap absorption, similar to previous reports [7, 19, 34]. A very weak absorbance feature is observed at about 2.55 eV, which we attribute to excitations from the few occurrences of NSs with high number of stacked layers (**Figure 1a** and **Figure S1a**) and bulk-like nanostructure's properties. The NSs exhibited broad PL with an emission peak position at 644.8 nm, a FWHM of 159.9 nm, and a red-shift with respect to the absorption peak of 207.0 nm. The broad emission spectrum and huge Stokes shift are considered to be a result of strong electron-phonon coupling and emission from self-trap excitons, previously widely discussed for different types of perovskites[35, 36]. The room temperature time-resolved PL decay was fitted to a tri-exponential function, which produced a fast and intermediate component with lifetimes of 5.9 and 163.2 ns, respectively, and a slow decay of 3.2 μs (**Figure 1h**). The relatively faster and intermediate decay processes are attributed to the recombination processes due to surface defects and trap states. The slow component is associated with radiative recombination processes linked to self-trapped exciton states, which are assumed to be easily depopulated by structural defects [12, 36].

## 2.2 Role of Temperature and Ligands

The role of the final reaction temperature in the colloidal synthesis of $Cs_2AgBiBr_6$ double perovskites on their structural and optical characteristics was investigated. **Figure 2a** shows the TEM image of the samples prepared at final reaction temperatures ranging from 200 to 290 °C. The samples prepared at 200 °C disclosed small nanoplatelets (10-50 nm) and few large sheets with lateral dimensions ranging from 50 to 500 nm. The lateral size of the nanoplatelets increases with the reaction temperature up to 250 °C, where nanosheets with a maximum lateral size of 1.4 μm were obtained. As mentioned before, this enhanced lateral growth is due to an increase in the thermal energy in the system at such a high temperature, which promotes the migration and growth of smaller nanoplatelets into larger sheets. At the final reaction temperature of 260 °C, the nanosheets began to deform and eventually produced bulk-like particles at 290 °C. This deformation is due to high reactivity at elevated temperatures that causes instability in the shelling of the nanoplatelets by the ligands, leading to inhomogeneous crystal growth, formation of byproducts, and eventual growth into bulk-like crystals. The XRD (**Figure S4a**) affirms the good quality of the crystals prepared at final reaction temperatures between 200-250 °C, as all the peaks observed clearly matched the cubic phase of $Cs_2AgBiBr_6$ double perovskites. Extra XRD peaks were observed for samples prepared at 260 °C and above, at 2θ values of 30.95, 44.34, and 50.08°, which are associated with the (200), (220), and (222) planes of the cubic crystal structure of $AgBr_3$, respectively (ICSD card number: 01-071-4692). Furthermore, the intensity of the XRD peaks along the z-direction increases significantly, especially for samples prepared at 270-290 °C, which affirms the growth of the NSs to bulk-like particles. The UV-Vis absorption spectra show confined excitonic excitation peaks around 2.83 eV (expected to be near the band edge) for samples prepared between 200-250 °C. The exciton absorption peaks were broadened and red-shifted by 0.06 and 0.11 eV for the samples prepared at 270 and 290 °C, respectively, and implies a change towards bulk-like nanostructures of $Cs_2AgBiBr_6$. Further extending the final reaction time beyond the normal 10 min (20-60 min) also resulted in the formation of byproducts and eventual growth towards bulk-like particles after 60 min, which is also due to unstable ligand shells around the platelets and excessive reactivity in the system (**Figure S5a and S5b**). The excitonic absorption peaks became broadened and red-shifted by 0.02 and 0.07eV when the final reaction time was increased to 20 and 60 min, respectively, which indicates a decrease in quantum confinement. All the samples prepared under different reaction temperature and time conditions exhibited broad PL emission at similar wavelengths of 644.8 ± 0.04 nm (**Figure S4c** and inset of **Figure S5c**).

Another crucial factor in the colloidal synthesis of nanostructures is the role ligands play in controlling their shape, size, uniformity, and structural defects. The role of key ligands – OA and OLAm, used in the synthesis of $Cs_2AgBiBr_6$, was studied by varying the concentration of each ligand while keeping other reaction parameters fixed. Optimal conditions for obtaining large NSs were observed at concentrations of 1.0-1.5 mL for OA and 1.0 mL for OLAm and a final reaction temperature of 250 °C. **Figure 2b** shows the TEM micrographs of $Cs_2AgBiBr_6$ samples prepared at various concentrations of OA. A lower concentration of OA (0.5 ml) largely produces small nanoplatelets of 5-80 nm size and some byproducts of $AgBr_3$, as seen in the XRD data (**Figure S6a**). The interaction between OA and the cations on the surface of $Cs_2AgBiBr_6$ nanocrystal is very weak, while OLAm binds strongly to the nanocrystals (NCs), as earlier reported by Zhang *et al.* in 2019. Zhang *et. al.* backed this with experimental evidence from nuclear magnetic resonance (NMR) and nuclear overhauser effect (NOE) spectroscopy studies [37]; however, the fundamental reason for the weak interaction of OA with the surface of the NCs has not yet been fully clarified. Notwithstanding, since OA and OLAm are both L-type ligands but with a carboxylic and amine headgroups, respectively [38], the pH of the nanocrystal's surface was probably not basic enough to allow for the deprotonation of OA to form NCs-OA bonds. An increase in pH has been previously linked to enhanced deprotonation of OA on material's surfaces [39, 40]. Nonetheless, the low amount of the OA ligand limits the acid-base equilibrium of the binary ligands needed to provide a stable ligand shell for the crystal growth of $Cs_2AgBiBr_6$ NSs. Further increase in OA concentration above 1.5 mL results in the formation of inhomogeneous NSs. The uniformity of the crystals and the formation of byproducts were highly sensitive to the amount of OLAm in the reaction system, since OLAm binds strongly to the surface of the nanocrystals. **Figure 2b** presents the TEM images of $Cs_2AgBiBr_6$ samples prepared at various concentrations of OLAm. The OLAm concentration of 1.0 ml favours the growth of $Cs_2AgBiBr_6$ NSs with no noticeable byproducts. Low concentration of OLAM led to a limited ligand shell and increased the reactivity in the system, leading to the growth of inhomogeneous sheets. Meanwhile, a further increase in OLAm concentration to 1.5 ml largely produces thick micro- and nanocrystals and a few nanosheets. Excess OLAm (2.0-2.5 ml) results in the formation of a densely packed ligand shell around the cations, which makes it difficult for complex structures to form and favours the formation of simpler compounds such as $AgBr_3$, $BiBr_3$, and $Cs_3Bi_2Br_9$ as byproducts (**Figure S7a**). So, while an increase or decrease in the ratio of OA/OLAm leads to the formation of byproducts, OA largely affects the lateral dimension of the nanocrystals, while OLAm significantly influences the uniformity of the NSs and their structural integrity. Interestingly, the excitonic absorption peak and PL emission positions were not affected by the variation of OA (**Figure S6b and S6c**) and OLAm (**Figure S7b and S7c**) concentrations in the reaction mixture: an indication of a relatively stable band structure that is not affected by morphological changes.

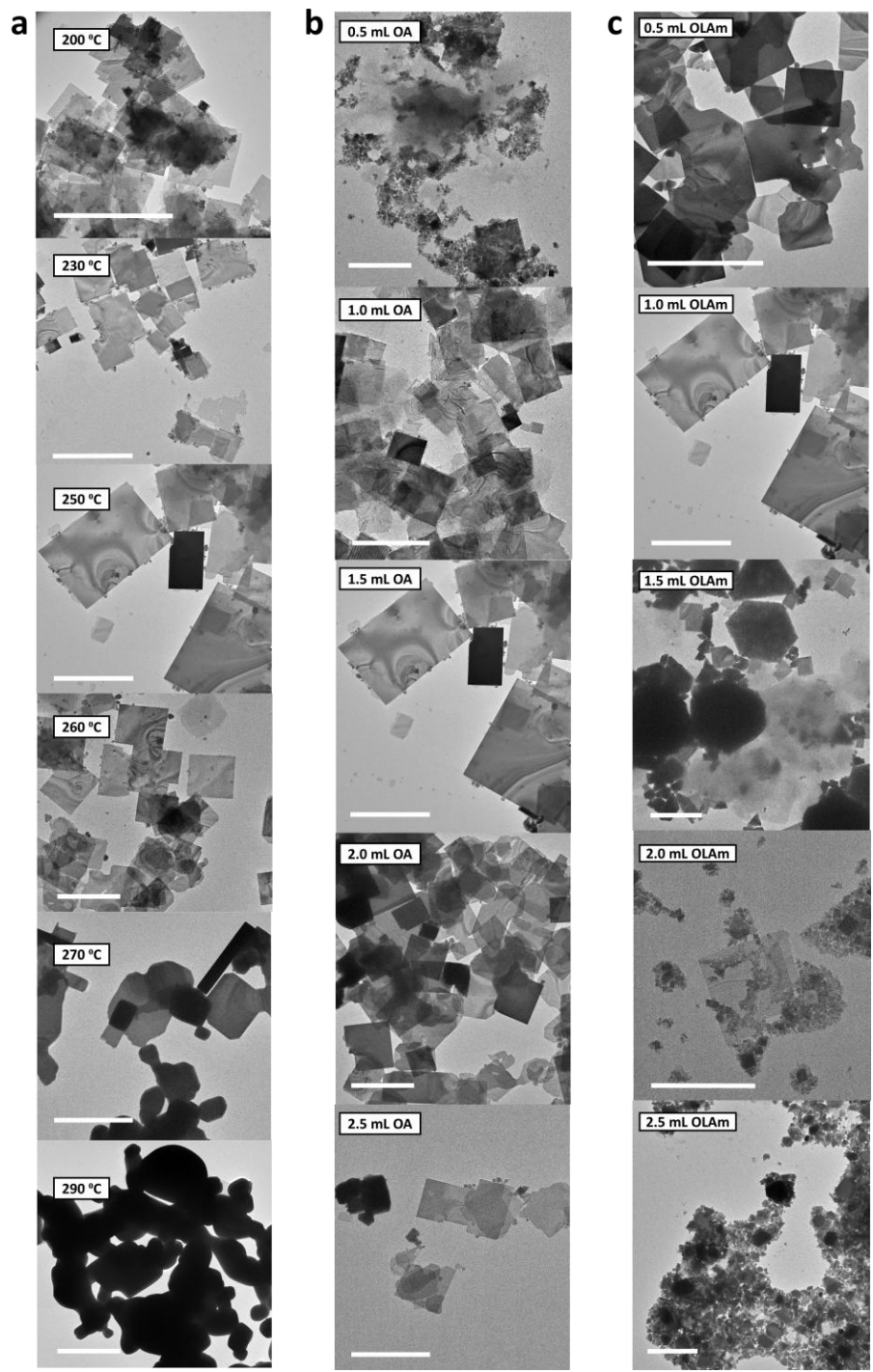

**Figure 2**. TEM images of colloidal $Cs_2AgBiBr_6$ samples prepared a) at different final reaction temperatures, b) using various concentrations of OA, and c) OLAm. The scale bars shown in the micrographs represent 1 μm.

## 2.3 Photodetection properties of $Cs_2AgBiBr_6$ nanosheets

Generally, photodetection involves the absorption of photons by a suitable material followed by optical transitions that produce free charge carriers, which are extracted at the electrodes with the aid of an electric field to yield photoinduced current. $Cs_2AgBiBr_6$ is a promising material for many applications in optoelectronic devices, especially where photon absorption and charge extraction are paramount, such as photodetection. Moreover, $Cs_2AgBiBr_6$ is Pb-free and has exhibited superior stability under ambient conditions, temperature, and exposure to light over other conventional halide hybrid organic-inorganic perovskites such as $MAPbBX_3$ (X=I, Br, or Cl) [41, 42]. In addition, they have shown desirable properties suitable for photodetection applications, such as a long carrier lifetime [12, 43], a low minority carrier diffusion length of 700 nm up to 2.44 μm (down to 123 K) [7], and high resistivity that results in low noise and reduced dark current [44].

A schematic representation of the fabricated quasi-2D $Cs_2AgBiBr_6$ NSs-based photodetection device is presented in the inset of **Figure 3(a)** with an expanded view given in **Figure S8a**. The device consists of the colloidal $Cs_2AgBiBr_6$ NSs deposited onto two interdigited Au electrodes, and the SEM micrograph, which shows all the regions on the device's surface, is given in **Figure S8b**. The device was annealed at 200 ºC for 10 min to improve ohmic contacts and remove insulating ligand barriers that can limit charge transport and device performance. The SEM image of the NSs before and after annealing is shown in **Figures S8c** and **S8d**, respectively. The micrographs disclosed $Cs_2AgBiBr_6$ NSs for the annealed device and unannealed sample (deposited on Si/SiO substrate), with no noticeable difference in shape and size. The UV-Vis absorption of the NSs that was annealed and redispersed into toluene also shows an absorption spectrum similar to the freshly prepared colloidal samples and comparable confined excitonic peak positions (**Figure S9a**). Furthermore, the XRD of the annealed and the unannealed colloidal NSs drop-casted onto glass substrates shows a similar XRD pattern for both samples (**Figure S9b**). The microstrain ($\varepsilon$) in the crystal lattice was evaluated from XRD data using the Williamson-Hall method, and similar values of 1.15 and $1.21 \times 10^{-3}$ were obtained for both annealed and unannealed $Cs_2AgBiBr_6$ NSs, respectively (**Figure S10a** and **S10b**). Therefore, optical and structural analysis suggest that the annealing of the photodetection device did not significantly alter the structural and optical features of the NS.

The current density-voltage curves of the $Cs_2AgBiBr_6$ NSs photodetection device in dark and under illumination conditions are presented in **Figure 3a**. A dark current density of $4.0 \times 10^{-8}$ $A/cm^2$ was observed for the photodetection device at a bias voltage of 5 V. This low dark current response is likely due to the charge injection barrier at the metal-semiconductor interface [45, 46]. The photocurrent notably increases under illumination and produces a remarkable on/off ratio of $2.39 \times 10^4$ at 17.7 $mW/cm^2$. This is higher than many reported values for $Cs_2AgBiBr_6$ [20, 46-48] and some Pb-based hybrid organic-inorganic perovskite photodetectors (**Table 1**). The on/off ratio of the photodetection device increases with increasing power density (**Figure S11a**) due to enhanced photogeneration of charge carriers. The high on/off ratio recorded for the photodetection device is attributed to the low dark current attained by the device, coupled with the high quality of the NSs, reduction of detrimental trap states at grain boundaries because of their large lateral sizes, and long charge carrier lifetimes, which are beneficial for charge extraction and high device performance.

The large lateral dimension of the synthesized NSs will bridge the electrode spacing of the device more efficiently, and limit defects due to grain boundaries and the consequent charge recombination to improve charge transport and photodetection performance [31]. More so, unlike many Pd-based organic-inorganic perovskites such as MAPbX$_3$ (Br, I, and Cl), the Cs$_2$AgBiBr$_6$ NSs could withstand high-temperature calcination (200 °C) without any significant distortion of their crystal structure, which was vital in improving charge transport and device performance. The dependence of photocurrent on incident power density is presented in **Figure 3b** and **Figure S11b**. The photocurrent increases with increasing power density of the incident photons because of the enhancement in the photogeneration of excitons. The photodetection device mildly treated at 50 °C for 30 min produces a photocurrent that is only $10^3$ higher than the dark current (**Figure S12a and S12b**): one order less than the value obtained for the annealed device at 200° C (**Figure 3a**). The reduced performance is mainly attributed to poor ohmic contacts at the Au/Cs$_2$AgBiBr$_6$ NSs interface and the presence of excess organic ligands that served as non-conducting barriers for charge transport. The annealing of Cs$_2$AgBiBr$_6$-based photodetectors at 75-285 °C has been previously reported, owing to the good thermal stability of Cs$_2$AgBiBr$_6$ [18, 34, 49, 50].

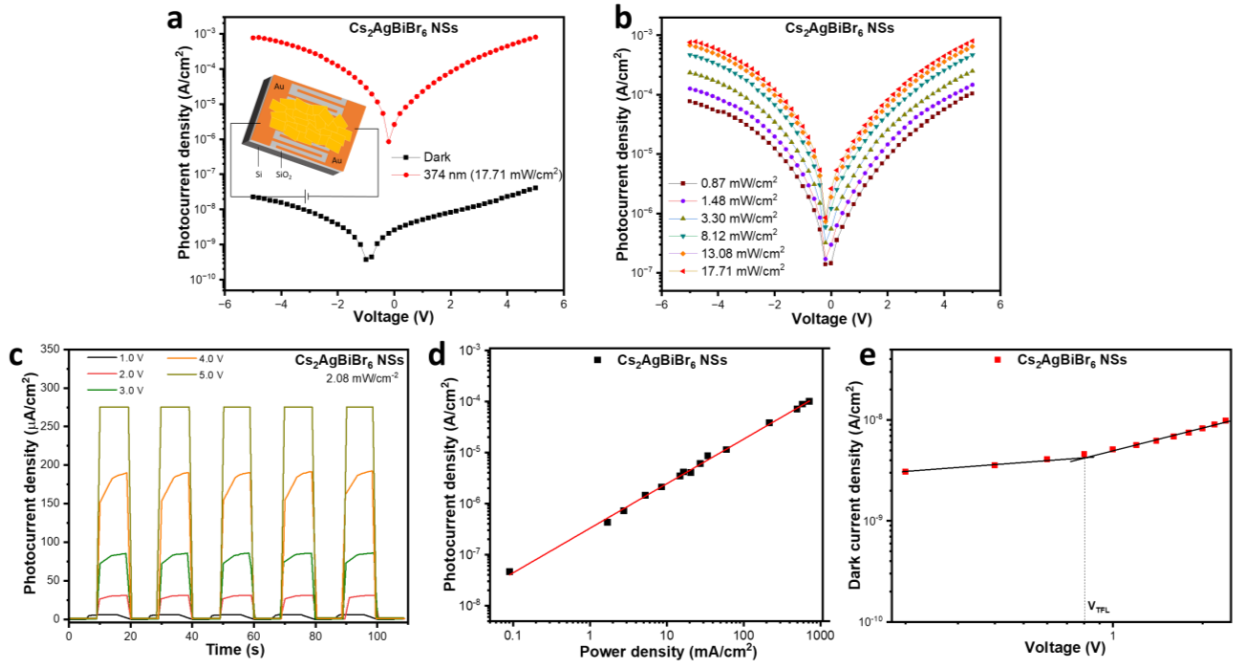

**Figure 3**. Current-voltage characteristics of Cs$_2$AgBiBr$_6$ NSs photodetector. a) current under dark and illumination conditions within a voltage bias range of -5.0 to 5.0 V: the inset shows the schematic diagram of the photodetection device, b) photocurrent at various incident light power densities and different voltages, c) transient photocurrent response at various voltages and fixed illumination power of 2.08 mW/cm$^2$, d) photocurrent density at different power densities obtained at 4.0 V and used to evaluate the LDR, and e) a plot of dark current density against different voltages.

Transient photocurrent density was also measured at different bias voltages for the Cs$_2$AgBiBr$_6$ NSs photodetection device at an incident power density of 2.08 mW/cm$^2$ and a signal frequency of 50 mHz (equivalent to a laser on/off duration of 20 s). The photocurrent density increases with increasing bias voltage **(Figure 3c)** due to the enhancement of electric field and carrier drift velocity that promote charge transport and extraction. In a similar measurement at a constant voltage of 2.0 V, the transient photocurrent also increases with increasing power density of the incident photons as expected **(Figure S13a)**. A linear dynamic range (LDR) was deduced for the $Cs_2AgBiBr_6$ NSs photodetectors from their photocurrent density measurements at various power densities, as given in **Figure 3d**. The LDR was evaluated using Equation 1[24]:

$$LDR = 20\log\left(\frac{I_{max}}{I_{min}}\right) \qquad 1$$

where $I_{max}$ and $I_{min}$ denote the maximum and minimum photocurrent within the linear range, respectively. The LDR evaluated for the device is 67 dB. Lower LDR values have been reported for some $Cs_2AgBiBr_6$-based photodetectors [51-53], as well as many transport-layer-free Pb-based organic-inorganic hybrid photodetectors [54-56]. The LDR defines the region within which the output of the photodetection device scales linearly with respect to the intensity of the incident light input: a necessary condition for sustaining a consistent responsivity [57]. The plot of photocurrent against incident power was fitted to the power law $I = AP^\theta$, where A is a constant and $\theta$ is a parameter that reflects the trapping and recombination processes of photogenerated carriers in the device ($\theta < 1$) [47]. After fitting the plot to the above relation, a $\theta$ value of 0.69 was obtained for the photodetection device **(Figure S13b)**. Fractional values of $\theta$ close to 1 are desirable as they indicate suppression of charge recombination processes in the device [45]. The $\theta$ value obtained for the $Cs_2AgBiBr_6$ NSs photodetector heated at 50 °C was 29.0 % less when compared to the device annealed at 200 °C during device fabrication **(Figure S13b and S13c)**. This is linked to the limitation of charge transport posed by poor ohmic contact and non-conducting organic ligands, which leads to high charge trapping rates and recombination. Furthermore, to evaluate the trap density, $N_t$, of the $Cs_2AgBiBr_6$ NSs photodetector, the space charge limited current (SCLC) analysis was performed, and the dark current density is plotted against various bias voltages to obtain the value of the trap filling limit (TFL) voltage, $V_{TFL}$, as shown in **Figure 3e**. A linear response is seen in the ohmic region at lower voltages, as indicated in the plot. At a higher voltage, a deviation from this ohmic behavior is observed at the trap filling limit, where a $V_{TFL}$ value of 0.805 V was obtained for the device. At the TFL, the trap states are occupied by charge carriers. So, the trap density can be evaluated using Equation 2:

$$N_t = \frac{2\varepsilon_r\varepsilon_o V_{TFL}}{eL^2} \qquad 2$$

where the relative dielectric constant is represented by $\varepsilon_r = 51$ [32], $\varepsilon_o$ denotes the vacuum permittivity, $e$ is the electronic charge, and the channel length, $L = 5$ μm. The $N_t$ of $1.82 \times 10^{14}$ cm$^{-3}$ was evaluated for the photodetection device. Limiting trap states is paramount for reducing charge recombination rates and improving carrier extraction to boost photodetection performance.

The 3-dB bandwidth or cut-off frequency, which characterizes the device's speed, was obtained from the frequency responses measured at laser pulsing frequencies of 0.05 to 2500 Hz. The normalized relative photocurrent ($I_{max}$-$I_{min}$) obtained at different frequencies was plotted as shown in **Figure 4a** and used to deduce the 3dB frequency of photodetectors at $(I_{max}-I_{min})/\sqrt{2}$, which equates to 70.7% of the maximum $I_{max}-I_{min}$ value (**Figure 4a**). A 3dB bandwidth of 363 Hz was obtained for the device. The response times of the $Cs_2AgBiBr_6$ NSs were evaluated at 400 Hz (near the 3dB frequency) and yield fast rise and fall times of 857 and 829 µs, respectively (**Figure 4b**). Despite the relatively large values related to the intrinsically long lifetimes of the charge carriers, these response times outperformed many pristine and modified $Cs_2AgBiBr_6$-based photodetectors and some conventional organic-inorganic Pb-based perovskite photodetection devices, as outlined in **Table 1**. The fast response times obtained for the device can be attributed to the good quality of colloidal quasi-2D nanocrystals and the large lateral size of the NSs that effectively bridged the electrode spacing and limits traps due to grain boundaries, which is beneficial for efficient charge transport.

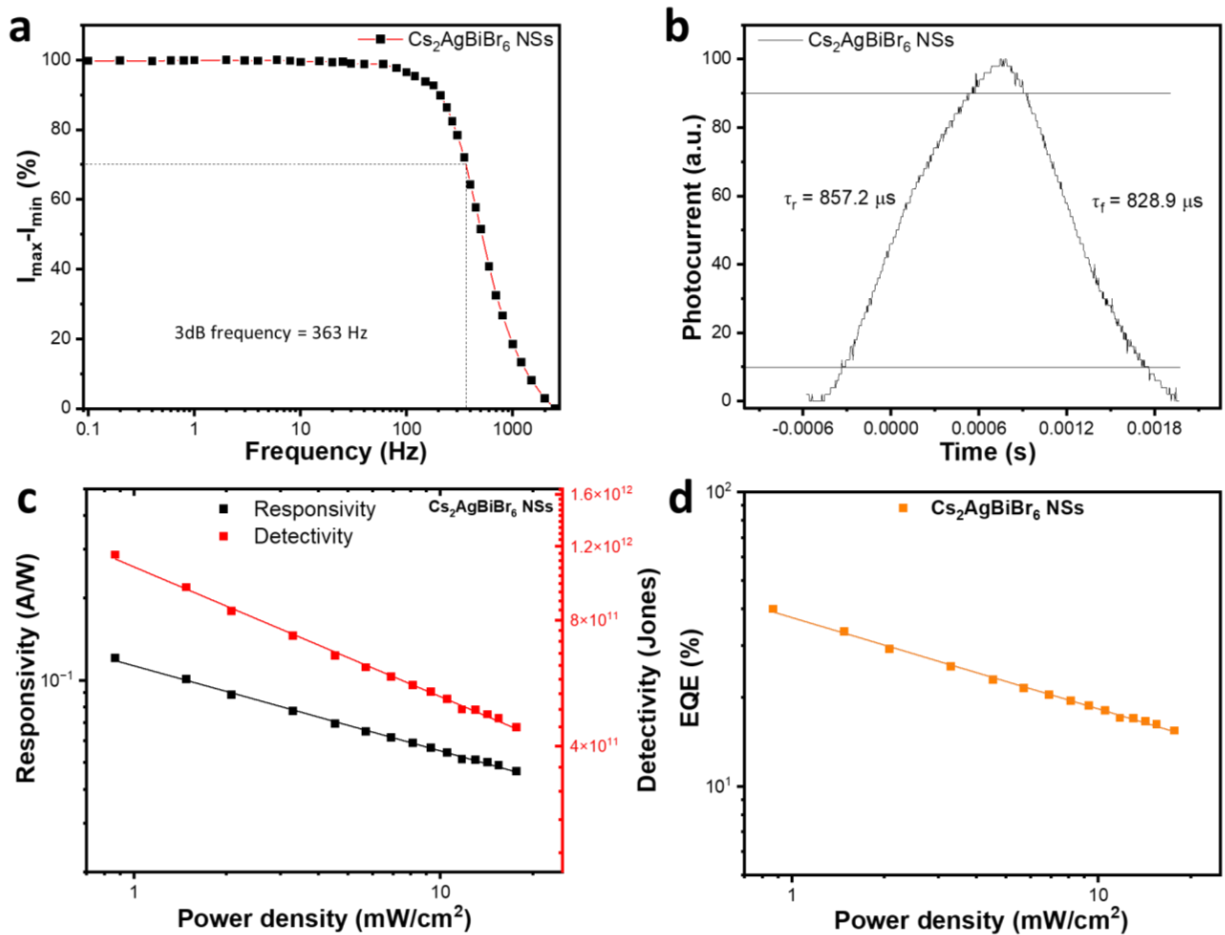

**Figure 4**. $Cs_2AgBiBr_6$ NSs photodetector. a) frequency responses obtained under 4.0 V bias voltage and 2.08 mW/cm² illumination pulse power, b) rise and fall times at 400 Hz, c) responsivity, detectivity at different power densities, and d) external quantum efficiency (EQE), evaluated at 5.0 V bias.

The photodetection performance of the Cs$_2$AgBiBr$_6$ NSs was further assessed by evaluating their responsivity (*R*), detectivity (*D*), and external quantum efficiency (*EQE*) using the relations in Equations 3, 4, and 5 [24]:

$$R = \frac{I_{photo} - I_{dark}}{PA} \quad \quad 3$$

$$D = \frac{Rhc}{e\lambda} \quad \quad 4$$

$$EQE = \frac{R}{\sqrt{e(I_{dark}/A)}} \quad \quad 5$$

where $I_{photo}$ and $I_{dark}$ are current under illumination and dark conditions, respectively, while *P* denotes the power density of incident light, *A* represents the active area of the photodetector, and *h*, *c*, *e*, and $\lambda$ denote the Planck constant, speed of light, electron charge, and the wavelength of the laser, respectively. The responsivity and detectivity of the device was evaluated at 5.0 V, and different power densities are shown in **Figure 4c**, while the *EQE* plot is presented in **Figure 4d**. The responsivity, detectivity, and *EQE* were observed to decrease with increasing laser power density. This behavior is similar to other observations [24, 58] and is associated with the prevalence of bimolecular recombination of charge carriers at higher power [58]. The NSs yield a high detectivity value of 1.15 × 10$^{12}$ Jones, outperforming many reported values for pristine and doped Cs$_2$AgBiBr$_6$ photodetectors, devices that engage heterojunctions and charge transport layers in their architecture, and conventional organic-inorganic halide perovskites (**Table 1**). The detectivity of the photodetector defines its capacity to detect weak optical signals and is a vital parameter in determining device performance. The responsivity, which quantifies the photocurrent per unit area for a given optical power, was evaluated and yields 121 mA/W. An *EQE* of 40% was deduced for the device, which represents the ratio of the number of photoexcited charge carriers extracted to produce photocurrent to the number of incident photons.

**Table 1**. Photodetection properties of pristine $Cs_2AgBiBr_6$ and $Cs_2AgBiBr_6$-based devices that consist of heterostructures and/or engaged transport layers in their architectures, and conventional organic-inorganic halide Pd-based perovskite photodetectors.

| Type of perovskites | Material | Voltage (V), source wavelength (nm) | D (Jones) | R (A/W) | Rise/fall time (ms) | ON/OFF ratio | Ref. |
|---|---|---|---|---|---|---|---|
| Pristine $Cs_2AgBiBr_6$ | $Cs_2AgBiBr_6$ | 5.0, 375 | $1.15 \times 10^{12}$ | 0.12 | 0.857 / 0.829 | $2.39 \times 10^4$ | This work |
| | $Cs_2AgBiBr_6$ | 0.0, white light, | $2.5 \times 10^7$ | 0.0001 | 1.3/9.0 | 40 (405 nm Laser) | [47] |
| | $Cs_2AgBiBr_6$ | 5.0, 400 | $1.38 \times 10^9$ | 0.09 | 159/85 | 42 | [48] |
| | $Cs_2AgBiBr_6$ | 5.0, 530 | $4.63 \times 10^{12}$ | 12.1 | 110/160 | $3.5 \times 10^2$ | [20] |
| | $Cs_2AgBiBr_6$ | 5.0, 520 | $5.66 \times 10^{11}$ | 7.01 | 0.955/0.995 | $2.16 \times 10^4$ | [19] |
| | $Cs_2AgBiBr_6$ | 6.0, 530 | $1.60 \times 10^{13}$ | 15 | ~770 | - | [59] |
| | $Cs_2AgBiBr_6$ | -5.0, 530 | $1.1 \times 10^9$ | 0.00096 | 192/271 | 17.7 | [45] |
| $Cs_2AgBiBr_6$-based (heterostructures and/or engaged transport layers) | $TiO_2/Cs_2AgBiBr_6$ | 0.0, AM 1.5G | $5.1 \times 10^{11}$ | 0.05 | 160/100 | $2.57 \times 10^3$ | [18] |
| | $CsAgBr_2/Cs_2AgBiBr_6$ | 0.5, 440 | $4.6 \times 10^{10}$ | 0.06 | 1.4 /1.6 | - | [60] |
| | $Cs_2AgBiBr_6/SnO_2$ | 0.0, 350 | $2.1 \times 10^{10}$ | 0.11 | $< 3 \times 10^3$ | - | [44] |
| | $NiO:Mg/Cs_2AgBiBr_6/C_{60}/BCP$ | 0.0, - | $4.7 \times 10^{12}$ | 0.055 | 382/120 (465 nm laser) | - | [61] |
| | $PEDOT:PSS/TiO_2/Cs_2AgBiBr_6$ | -0.3, 405 | $9.29 \times 10^{12}$ | 0.666 | 166/163 | $1.23 \times 10^4$ | [49] |
| | $Cs_2AgBiBr_6/TiO_2$ | 0.0, 405 | $3.3 \times 10^{11}$ | - | 2.2/2.7 | - | [62] |
| organic-inorganic | $MAPbBr_3$-$PdSe_2$ | 2.0, 520, | $5.2 \times 10^{11}$ | 0.028 | 0.024/0.025 | - | [63] |
| | $Ni(OH)_2 \cdot 0.75H_2O$-$MAPbBr_3$ | -, 365 | $2.05 \times 10^{10}$ | 0.0019 | 84/76 (532 nm laser) | 131.2 | [64] |
| | $MAPbBrCl_2$ | 0.6, 440 | $1.41 \times 10^{11}$ | 0.051 | 340/920 | - | [55] |
| | $MAPbX_3(X=Br, I)$ | 0.0, 473 nm | $7.01 \times 10^{11}$ | 0.26 | 80/580 | $2 \times 10^5$ | [65] |
| | $MAPbBr_3$-$TiO_2$ | Solar Simulator | $\sim 10^{12}$ | 85 | 90/110 | $6.3 \times 10^2$ | [66] |

The energy band positions and alignment of the $Cs_2AgBiBr_6$ NSs and the Au electrodes in their equilibrium states before contact are given in **Figure 5a**. When in contact, electrons will flow from the semiconductor to the metal to attain equilibrium and a common Fermi level, since the work function of Au (5.1 eV) [46] is higher than that of $Cs_2AgBiBr_6$ based on values given in different reports [45, 48, 50]. This will cause energy band bending at both ends of the Au-$Cs_2AgBiBr_6$ interfaces, as illustrated in **Figure 5b**. Under dark conditions, charge flow is limited by the energy barrier at the Au-$Cs_2AgBiBr_6$ interfaces, which contributes to the low dark current observed for the photodetector. Under illumination, photogenerated charge carriers are excited to the conduction band (CB) to form electron-hole pairs. The electrons in the CB are driven towards the anode while the holes move in the opposite direction. The electric field that results from the external bias will aid the flow of charge carriers and boost the photocurrent generated by the device. Furthermore, since the photodetector is made of stacked layers of interconnected nanosheets, at room temperature, the charge carriers will move from one sheet to the other through both hopping and tunneling mechanisms, and get extracted at the electrodes. The grain boundaries would create defects that serve as trap centers for charge carriers, which will limit charge transport [31]. Hence, the large lateral size observed for the NSs helps to limit grain boundaries and contributes to the high photodetection performance recorded in the presented structures.

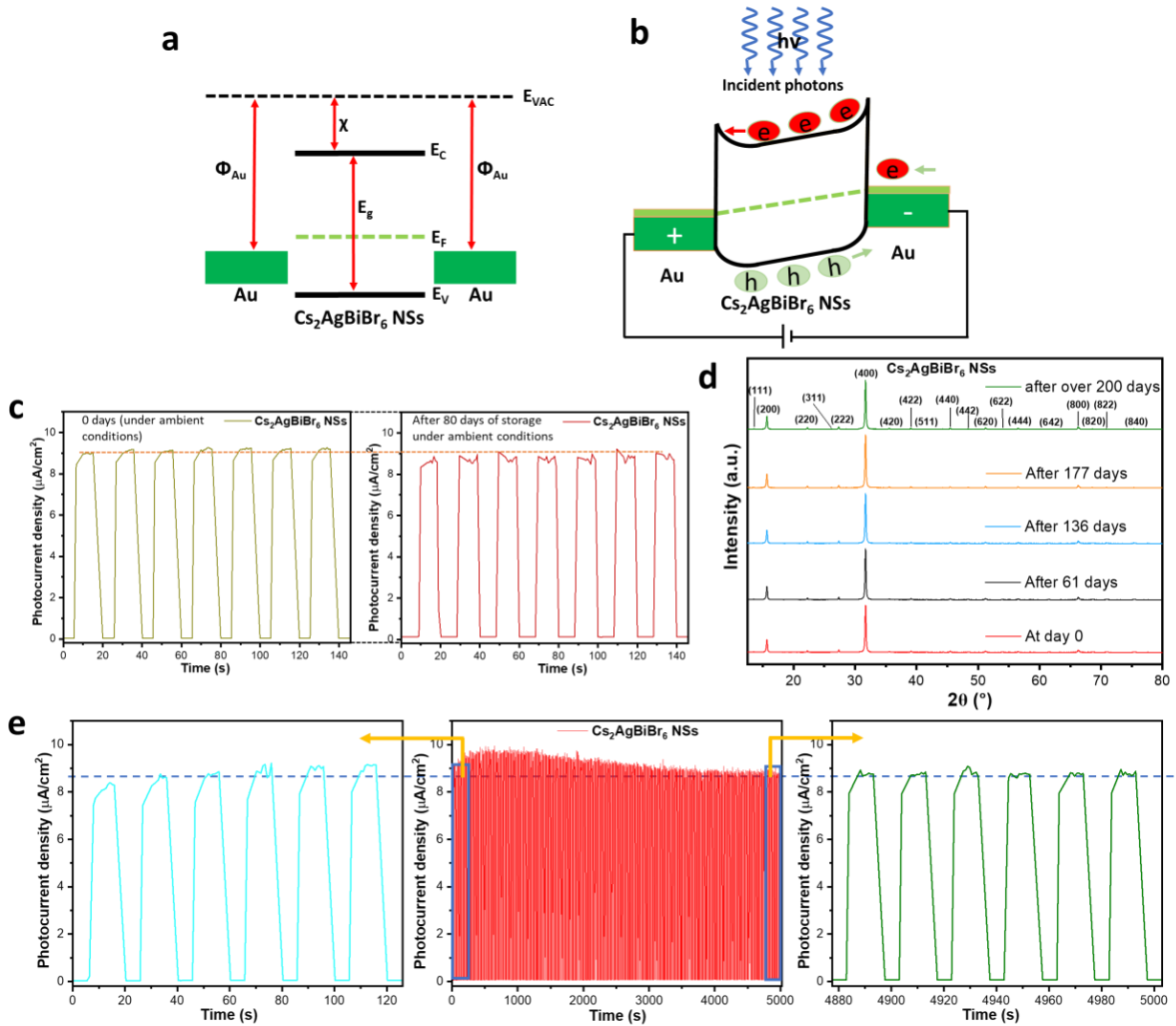

**Figure 5.** The energy band positions of $Cs_2AgBiBr_6$ NSs and Au electrodes a) before contact in their equilibrium states and b) after contact and illumination with incident photons, c) transient photocurrent of $Cs_2AgBiBr_6$ NSs photodetector at 0.87 mW/cm$^2$ illumination power and 3.0 V obtained under ambient conditions at day 0 and after storage for 80 days, and d) XRD patterns of the $Cs_2AgBiBr_6$ NSs between day 0 and after over 200 days of storage in ambient conditions, and e) the long-term time-resolved photocurrent analysis of the photodetection device.

The stability of the $Cs_2AgBiBr_6$ NSs photodetectors was investigated under ambient conditions. Transient photocurrent was recorded for the device and repeated after storing the sample under ambient conditions. The device demonstrated excellent stability, showing no noticeable difference in the on/off photocurrent after being kept under ambient conditions for 80 days (**Figure 5c**). Meanwhile, Pb-based organic-inorganic perovskite, $MAPbI_3$, photodetector, has been shown to lose 16 % of its photocurrent after storage in air for 30 days [65]. The stability of the NSs is further affirmed by the structural stability of the material when kept under ambient conditions for over 200 days. The XRD patterns obtain from intermittent measurements performed on the NSs between day 0 and after over 200 days solely revealed peaks for the cubic phase of $Cs_2AgBiBr_6$ double perovskites, with similar intensities, and no additional peaks for AgBr or other impurities (**Figure 5d**). Furthermore, the $Cs_2AgBiBr_6$ NSs photodetection device exhibited great long-term photostability and sustained a relatively consistent photocurrent in a transient current measurement over 5000 s (**Figure 5e**). The long-term stability test shows a slight initial increase in photocurrent. This is assumed to be due to some photo-activation phenomena by high-energy photons that resulted in the healing of some detrimental surface defects [19]. Further investigation will be required to fully explain this phenomenon.

## 3.0    Conclusions

In summary, in this work, quasi-2D $Cs_2AgBiBr_6$ nanosheets (NSs) were developed using the low-temperature injection colloidal synthesis and used to fabricate high-performance photodetectors. The reaction temperature and the concentration of ligands (OA and OLAm) played significant roles in the shape, size, and structural integrity of the synthesized $Cs_2AgBiBr_6$ NSs. The tuning and optimization of the ligands in the reaction system allowed for the synthesis of $Cs_2AgBiBr_6$ NSs at a higher final reaction temperature (250 °C), which leads to the formation of larger sheets, without any structural distortion. $Cs_2AgBiBr_6$ NSs of up to 1.4 µm lateral sizes, which are only a few nm thick, were synthesized. The high-performance photodetectors fabricated using the $Cs_2AgBiBr_6$ NSs yielded a notably high detectivity and on/off ratio of $1.15 \times 10^{12}$ Jones and $2.39 \times 10^4$, respectively. The device also recorded responsivity of 121 mA/W, and a fast sub-millisecond rise and decay time of 857 and 829 µs, respectively. The performance of the $Cs_2AgBiBr_6$ NSs photodetector is remarkable, especially because the device is free of transport layers, which often complicates device design and reduces the cost effectiveness for practical applications. This work demonstrates the potential of micro-sized colloidal quasi-2D $Cs_2AgBiBr_6$ lead-free double perovskite NSs with desirable characteristics for the design of high-performance photodetectors and other optoelectronic applications.

## 4.0    Experimental

### 4.1    Chemicals

Silver nitrate ($AgNO_3$, 99.8%), bismuth bromide ($BiBr_3$, >98.0%), Cesium carbonate ($Cs_2CO_3$, 99.9%), 1-octadecene (ODE, technical grade 90%), oleic acid (OA, technical grade 90%), oleylamine (OLAm, technical grade 70%), hydrobromic acid (HBr, 47%), and toluene (from VWR, 99.5%). All chemicals were purchased from Sigma-Aldrich unless stated, and used as received without further purification.

## 4.2  Preparation of Cs-oleate solution

10 ml of OLAm was added to 825 mg of $Cs_2CO_3$ that was measured and loaded into a 50 mL 3-neck flask. In a Schlenk line, under continuous Ar flow, the mixture was heated to 120 °C and degassed by applying vacuum pressure for 30 min. The solution was further heated to 150 °C, kept at the same temperature for 1 hr under Ar flow, and allowed to naturally cool down to room temperature for subsequent use.

## 4.3  Synthesis of $Cs_2AgBiBr_6$

The procedure developed by Liu et al. for the preparation of $Cs_2AgBiX_6$ (X=Cl, Br, and I), nanoplatelets was used with some modifications to the final reaction temperature and the amount of OA used to prepare quasi-2D $Cs_2AgBiBr_6$ nanosheets. In brief, 0.2 mmol $AgNO_3$, 0.1 mmol $BiBr_3$, 4 mL ODE, 1.5 mL OA, 1.0 ml OLAm, and 0.1 mL HBr were loaded into a 25 mL three-neck flask, heated to 120 °C under argon flow, and degassed by applying vacuum for 30 mins to remove residual air and water. After degassing, the mixture was heated to 200 ⁰C for 10 min under Argon flow to fully dissolve all precursors and was allowed to cool naturally to room temperature (25 °C). At room temperature, 0.3 mL of the prepared Cs-oleate solution was injected into the mixture and kept for 10 min at room temperature while stirring. The mixture was heated to 250 °C and kept for 10 min at this temperature. Afterwards, the heating mantle was removed to quench the reaction and cooled to room temperature. The crude solution obtained was centrifuged at 8000 rpm for 5 min to precipitate the sample from the final reaction mixture. The sample was dispersed in toluene and centrifuged again under the same conditions as the crude solution. Lastly, the supernatant was decanted, and the $Cs_2AgBiBr_6$ NSs sample was re-dispersed in 7.5 mL toluene and stored for further use.

Several samples were also synthesized to investigate the role of temperature and ligands in the colloidal synthesis of $Cs_2AgBiBr_6$ double perovskites. Samples were prepared, following the same procedure described for the NSs synthesis above, but at final reaction temperatures of 210, 230, 260, 270, and 290 °C. Also, following the same procedure, additional samples were prepared at 250 °C and final reaction times of 30 and 60 mins. More so, the role of OA in the reaction was investigated. Additional samples were prepared using 0.5, 1.0, 2.0, and 2.5 mL of OA in the reaction system while maintaining 1.0 mL OLAm and other reaction parameters described for the nonosheets synthesis above. In another set of samples, the amount of OLAm in the reaction mixture was 0.5, 1.5, 2.0, and 2.5 mL, while 1.5 mL of OA and other parameters were maintained. All samples were cleaned following the same procedure described for the $Cs_2AgBiBr_6$ NSs samples and dispersed in toluene.

## 4.4 Device fabrication

The Si/SiO2 substrates were obtained from Addison Engineering, Inc. San Jose, CA (300 nm ± 5% thermal oxide and resistivity of 005-020 Ω-cm) with interdigitated gold electrodes and 5 μm electrode spacing was used for the fabrication of the photodetectors. The substrates were cleaned by sonicating in acetone for 2 min and dried with flowing nitrogen gas. 50 μL of the stored $Cs_2AgBiBr_6$ NSs sample was further dissolved in 50 μL of toluene. 20 μL of the resulting solution was drop-casted onto the Si/SiO$_2$ substrate (with interdigitated gold electrodes) and allowed to dry in air. This was repeated once to obtain thin films of $Cs_2AgBiBr_6$ NSs on the substrates. The devices were placed in a vacuum oven and purged twice with Ar to remove any residual air in the chamber. The temperature in the oven was increased to 200 °C at a heating rate of ~2 °C/min under flowing Ar gas, calcined for 10 min, and allowed to cool naturally to room temperature to obtain quasi-2D $Cs_2AgBiBr_6$ NSs-based photodetection devices.

## 4.5 Photodetection measurements

The fabricated devices were placed in a probe station and kept under vacuum for evaluation of photodetection properties. The I–V characteristics of the $Cs_2AgBiBr_6$ NSs photodetection devices were studied by employing the Keithley 4200 semiconductor characterization equipment, with measurements conducted under dark and various incident light power density and voltage bias conditions. The 374 nm excitation laser served as the main incident light source. A 405 nm excitation laser was used to measure I-V responses under different illumination power densities and bias voltages to study the region within which the photocurrent scales linearly with respect to the intensity of the incident light (the laser provides much higher power density). Transient current responses were also measured for the photodetection device under various illumination power densities and bias voltages. The RIGOL function generator (DG4062) and the illumination for the laser (374 nm), which was converted into a pulsed light, were used in evaluating the device's frequency/time response. Using a power density of 2.08 mW/cm$^2$ from the pulse laser, a Tektronix oscilloscope (TDS2014B), and a bias voltage of 4 V (from SRS70, power source), the current-frequency response of the photodetection device was measured. The device's active area is 0.0003632 cm$^2$.

## 4.6 Characterizations

The prepared samples were further diluted in toluene before drop-casting onto transmission electron microscopy (TEM) copper grids and used to conduct TEM measurements using a Talos-L120C and EM-912 Omega at 120 kV. High-resolution TEM (HR-TEM) was conducted on $Cs_2AgBiBr_6$ NSs at 200 KV in STEM mode using a Jeol JEM-ARM200F NeoARM (probe-corrected) instrument.

A Zeiss EVO/MA10 instrument was used to conduct scanning electron microscopy (SEM) measurements on $Cs_2AgBiBr_6$ NSs deposited on bare Si/SiO$_2$ and the interdigitated gold electrode substrates.

X-ray diffraction (XRD) patterns of the prepared samples were measured using a Panalytical Aeris diffractometer with a Bragg-Brentano geometry and a copper anode with an X-ray wavelength of 0.154 nm from the Cu-k$\alpha$1 line.

An AFM (Park Systems XE-100) instrument in non-contact mode (ARROW-NCR-20, force constant $k$ = 42Nm−1, resonance frequency at 285 kHz) was used to measure the surface topography of the samples, which were prepared by drop-casting the diluted samples on Si/SiO$_2$ substrates.

The UV-Vis absorption spectra of the samples were obtained with a Lambda 1050+ spectrophotometer from Perkin Elmer equipped with an integration sphere.

A spectrofluorometer FS5 fluorescence spectrometer from Edinburgh Instruments was used to obtain the PL spectra of the samples. The time-resolved PL measurements were conducted using a picosecond laser attached to the spectrometer with an excitation wavelength of 375 nm and a repetition rate of 100 kHz. The decay profiles are tail-fitted with a bi-exponential function.

**Acknowledgements**

P.I.K. acknowledges Alexander von Humboldt-Stiftung/Foundation for the postdoctoral research fellowship. Deutsche Forschungsgemeinschaft (DFG, German Research Foundation) is acknowledged for funding of SFB 1477 "Light-Matter Interactions at Interfaces", project number 441234705, W03 and W05. C. K. also acknowledges the European Regional Development Fund of the European Union for funding the PL spectrometer (GHS-20-0035/P000376218) and X-ray diffractometer (GHS-20-0036/P000379642) and the DFG for funding an electron microscope Jeol NeoARM TEM (INST 264/161-1 FUGG) and an electron microscope Thermo Fisher Talos L120C (INST 264/188-1 FUGG).